\documentclass[12pt]{article}
\usepackage{amssymb,amsmath}
\begin{document}

\title{\textbf{Integrable cosmological models in the Einstein and in the Jordan frames and
Bianchi--I cosmology}}

\author{A.Yu.~Kamenshchik$^{a,b,}$\footnote{E-mail: Alexander.Kamenshchik@bo.infn.it},  E.O.~Pozdeeva$^{c,}$\footnote{E-mail: pozdeeva@www-hep.sinp.msu.ru}, \ A.~Tronconi$^{a,}$\footnote{E-mail: tronconi@bo.infn.it},\\ G.~Venturi$^{a,}$\footnote{E-mail: giovanni.venturi@bo.infn.it}, S.Yu.~Vernov$^{c,}$\footnote{E-mail: svernov@theory.sinp.msu.ru}\\
\small $^a$ Dipartimento di Fisica e Astronomia and INFN, \small  Via Irnerio 46, 40126, Bologna,
Italy\\
\small $^b$ L.D. Landau Institute for Theoretical Physics of  \small  the Russian
Academy of Sciences,\\ \small  Kosygin str.~2, 119334, Moscow, Russia\\
\small $^c$ Skobeltsyn Institute of Nuclear Physics, Lomonosov Moscow State University,\\ \small  Leninskie Gory~1, 119991, Moscow, Russia}

\date{ \ }

\maketitle

\begin{abstract}
We study integrable models in the Bianchi I metric case with scalar fields minimally and non-minimally
coupled with gravity and the correspondence between their general solutions. Using the model with a minimally coupled scalar field and a constant potential as an example, we demonstrate how to obtain the general solutions of the corresponding models in the Jordan frame.
\end{abstract}
\vspace*{6pt}


\section{Introduction}

Cosmological models with scalar fields play a central role in the description of the global evolution of the Universe. Models with the Ricci scalar multiplied by a function of the scalar field are quite natural because quantum corrections to the effective action with minimal coupling include non-minimal coupling terms~\cite{ChernikovTagirov,Callan:1970ze}. Modern inflationary models with a non-minimally coupled scalar field not only do not contradict the recent observational data~\cite{Planck2015}, but also connect cosmology and particle physics~\cite{HiggsInflation,nonmin-infl,EOPV2014}.

We consider a cosmological model with the following action
\begin{equation}
S =\int d^4x\sqrt{-g}\left[U(\sigma)R - \frac12g^{\mu\nu}\sigma_{,\mu}\sigma_{,\nu}-V(\sigma)\right],
\label{action}
\end{equation}
where $U(\sigma)$ and $V(\sigma)$ are differentiable  functions of the scalar field~$\sigma$.

In our previous papers~\cite{KPTVV2013,KPTVV2015,KPTVV2016} we considered integrable cosmological models with a Friedmann--Lema\^{i}tre--Robertson--Walker (FLRW) metric and found new integrable models with a non-minimal coupling by using the knowledge of
the corresponding minimally coupled models. The goal of this paper is to sketch the generalization of the method
developed before to the case of the Bianchi--I cosmology.

\section{The model with non-minimal coupling in the Bianchi I metric}
\label{Sec2}

Let us consider the Bianchi I metric with the interval
\begin{equation}
ds^2 = {}-N^2(\tau)d\tau^2 +a^2(\tau)\left(e^{2\beta_1(\tau)}dx_1^2+e^{2\beta_2(\tau)}dx_2^2+e^{2\beta_3(\tau)}dx_3^2\right),
\label{Biametr}
\end{equation}
where $a(\tau)$ is the scale factor, $N(\tau)$ is the lapse function, and the functions $\beta_i(\tau)$ satisfy the constraint
$\beta_1(\tau)+\beta_2(\tau)+\beta_3(\tau)=0$.
Following the papers~\cite{Pereira,ABJV2009}, we introduce the shear
\begin{equation}
\theta\equiv \dot\beta_1^2+\dot\beta_2^2+\dot\beta_3^2=2\left(\dot\beta_1^2+\dot\beta_2^2+\dot\beta_1\dot\beta_2\right),
\end{equation}
here and hereafter a ``dot'' means a derivative with respect to time, whereas a ``prime'' means a derivative with respect to~$\sigma$.

On varying the action~(\ref{action}), one gets the following equations in the Bianchi I metric:
\begin{equation}\label{Equ00B1}
   \left(6h^2-\theta\right)U+6hU'\dot\sigma=\frac12\dot\sigma^2+N^2V\,,
\end{equation}
\begin{equation}
\begin{split}
 &4U\dot{h}+6Uh^2-4Uh\frac{\dot{N}}{N}+2U''\dot{\sigma}^2+U\left[\theta-2\ddot\beta_i-6h\dot\beta_i+2\frac{\dot N}{N}\dot\beta_i\right]+
 \\
&+2U'\left[\ddot{\sigma}+2h\dot{\sigma}-\dot\beta_i\dot{\sigma}-\dot{\sigma}\frac{\dot{N}}{N}\right]= {}-\frac12\dot{\sigma}^2+N^2V\,,
\end{split}
\label{Bianchi2}
\end{equation}
\begin{equation}
\ddot{\sigma}+\left(3h-\frac{\dot{N}}{N}\right)\dot{\sigma} -6U'\left[\dot{h}+2h^2-h\frac{\dot{N}}{N}+\frac16\theta\right]+N^2V' = 0\,,
\label{KGB1}
\end{equation}
where $h\equiv \dot a/a$.
We also get the equation for $\theta$ which can be integrated  easily:
\begin{equation}\label{Equtheta}
\dot \theta=2\left[\frac{\dot N}{N}-3h-\frac{\dot U}{U}\right]\theta \qquad \Rightarrow\qquad \theta=\frac{N^2}{U^2a^6}\theta_0.
\end{equation}
By definition $\theta\geqslant 0$, thereby, a constant $\theta_0\geqslant 0$.

\section{Integrable models with a minimal and non-minimal coupling}

Let us make a conformal transformation of the metric
$g_{\mu\nu} = \frac{U_0}{U}\tilde{g}_{\mu\nu}$,
where $U_0$ is a positive constant.
We also introduce  such a new scalar field $\phi$  that
\begin{equation}
\frac{d\phi}{d\sigma} = \frac{\sqrt{U_0(U+3U'^2)}}{U}
\quad\Rightarrow\quad
\phi = \int \frac{\sqrt{U_0(U+3U'^2)}}{U} d\sigma.
\label{scal1}
\end{equation}
As a result the action (\ref{action}) transforms to the following action with the minimal coupling:
\begin{equation}
S =\int d^4x\sqrt{-\tilde{g}}\left[U_0R(\tilde{g}) - \frac12\tilde{g}^{\mu\nu}\phi_{,\mu}\phi_{,\nu}-W(\phi)\right] ,\quad \mbox{where}\quad  W(\phi) = \frac{U_0^2 V(\sigma(\phi))}{U^2(\sigma(\phi))}.
\label{action1}
\end{equation}

In the Einstein frame the metric~(\ref{Biametr}) transforms to the following Bianchi I metric

\begin{equation}
\label{EinFreimMetric}
ds^2 ={}- \tilde{N}^2(\tau)d\tau^2 + \tilde{a}^2(\tau)\left(e^{2\beta_1(\tau)}dx_1^2+e^{2\beta_2(\tau)}dx_2^2+e^{2\beta_3(\tau)}dx_3^2\right),
\end{equation}
where the new lapse function and  the new scalar factor are
\begin{equation*}
\tilde{N} = \sqrt{\frac{U}{U_0}}N,\qquad \tilde{a} =  \sqrt{\frac{U}{U_0}}a.
\end{equation*}

The functions $\beta_i$ are the same in both frames.
The equations in the Einstein frame are:
\begin{equation}
U_0\left(6\tilde{h}^2-\theta\right)=\frac12\dot{\phi}^2+\tilde{N}^2W,\label{A8}
\end{equation}
\begin{equation}
4U_0\dot{\tilde{h}}+6U_0\tilde{h}^2-4U_0\tilde{h}\frac{\dot{\tilde{N}}}{\tilde{N}}+U_0\theta
= -\frac12\dot{\phi}^2+\tilde{N}^2W,
\label{A10}
\end{equation}
\begin{equation}
\ddot{\phi}+\left(3\tilde{h}-\frac{{\dot{\tilde{N}}}}{\tilde{N}}\right)\dot{\phi}
+\tilde{N}^2W_{,\phi} = 0,
\label{A7}
\end{equation}
\begin{equation}\label{EquthetaEinfr}
\dot \theta=2\left[\frac{\dot{\tilde{N}}}{\tilde{N}}-3\tilde{h}\right]\theta,
\end{equation}
where $\tilde{h} \equiv{\dot{\tilde{a}}}/{\tilde{a}}$.
It is easy to solve Eq.~(\ref{EquthetaEinfr}) and obtain
\begin{equation}
\label{theta1}
\theta=\theta_0\frac{\tilde{N}^2}{\tilde{a}^6U_0^2}=\theta_0\frac{{N}^2}{{a}^6U^2}.
\end{equation}

Let us suppose that for some potential $W$ we know the general solution of the system (\ref{A8})--(\ref{A7}), that is we know explicitly or in quadratures the functions $\phi(\tau)$, \ $\tilde{a}(\tau)$, \ $\tilde{N}(\tau)$. On using (\ref{theta1}), we obtain the function $\theta(\tau)$. We also suppose that the function~$\sigma(\phi)$ is known.
For this case, the general solution of the system of equations (\ref{Equ00B1})--(\ref{KGB1}) with the potential
$V(\sigma) =U^2(\sigma)W(\phi(\sigma))/{U_0^2}\,$,
is given by
\begin{equation}\label{GeNSol}
\sigma(\tau)=\sigma(\phi(\tau)),\quad a(\tau) = \sqrt{\frac{U_0}{U(\sigma(\phi(\tau)))}}\tilde{a}(\tau),\quad N(\tau) = \sqrt{\frac{U_0}{U(\sigma(\phi(\tau))}}\tilde{N}(\tau).
\end{equation}

Let us consider as an example the case of a constant potential: $W(\phi)=\Lambda>0$. On summing Eqs.~(\ref{A8}) and (\ref{A10}) and choosing $\tilde{N}=1$, we get the equation on the Hubble parameter that gives two solutions:
\begin{equation*}
%\label{hequ}
 \dot{\tilde{h}}+3\tilde{h}^2=\frac{\Lambda}{2U_0} \quad \Rightarrow \quad
 \tilde{h}_1=\sqrt{\frac{\Lambda}{6U_0}}\tanh(\upsilon),\, \tilde{h}_2=\sqrt{\frac{\Lambda}{6U_0}}\coth(\upsilon),\  \upsilon=\sqrt{\frac{3\Lambda}{2U_0}}(\tau-\tau_0),
\end{equation*}
These solutions correspond to scale factors and, on solving Eq.~(\ref{A7}), $\dot\phi$:
\begin{equation*}
\tilde{a}_1=\tilde{a}_0\cosh(\upsilon)^{1/3},\quad \tilde{a}_2=\tilde{A}_0\sinh(\upsilon)^{1/3},\qquad \dot{\phi}_1=\frac{c_1}{\cosh(\upsilon)},\quad  \dot{\phi}_2= \frac{c_2}{\sinh(\upsilon)}.
\end{equation*}
Here $\tau_0$, $\tilde{a}_0$, $\tilde{A}_0$, $c_1$ and $c_2$ are integration constants.
The functions obtained should satisfy Eq.~(\ref{A8}). After substitution we get
$\theta_0=-\tilde{a}_0^6U_0(\Lambda+c_1^2/2)$ for the first solution. The inequality $\theta_0<0$ means that such a solution does not exist and one should select the second solution. For the second solution Eq.~(\ref{A8}) gives
$\theta_0=\tilde{A}_0^6U_0(\Lambda-c_2^2/2)$. The function~$\phi_2$~is  equal to
\begin{equation}
\phi_2(\tau)={}-\frac{2c_2\sqrt{6U_0}}{3\sqrt{\Lambda}}\mathrm{arctanh}\left(e^{\upsilon}\right)+c_0,
\end{equation}
where $c_0$ is a constant. The condition $\theta_0\geqslant0$ gives $c_2^2\leqslant2\Lambda$.

Similar solutions have been found in the FLRW metric case~\cite{KPTVV2016,Aref'eva:2007yr}. The only difference from the FLRW case is the form of the constraint on the possible values of the integration constants that are defined by Eq.~(\ref{A8}). The general solution obtained for the model with a minimal coupling and a constant potential and formulae~(\ref{GeNSol}) allows one  to find general solutions for a model with a nonminimal coupling and the potential $V=\Lambda U^2/U_0^2$. Thus, we have a set of integrable cosmological models in the Bianchi I metric.

\section{Conclusions}

In this short paper we have shown that the method for the construction of
integrable models with a non-minimally coupled scalar field proposed in~\cite{KPTVV2013}, can be generalized to
Bianchi I models. We just considered a simple example with a constant potential, and we hope to present examples with more complicated potentials in a future publication~\cite{KPTVV2016B}. The integrable Bianchi--I model without a scalar field, but with constant cosmological, stiff matter and dust has been considered in papers~\cite{Khalatnikov:2003ph,Kamenshchik:2009dt}, where general solutions have been found. It would be interesting to generalize our approach to cosmological models with dust and radiation. The number of known integrable Bianchi I models is less than number of integrable FLRW models\footnote{The list of integrable FLRW models with a minimal coupling is presented in~\cite{Fre}, some integrable Bianchi I models are presented in~\cite{IntBianchi}.}. We hope that the proposed method can help one to obtain new integrable models with minimal and non-minimal coupling.

A.K. was partially supported by the RFBR grant 14-02-00894.
Research of E.P. is supported in part by grant MK-7835.2016.2  of the President of Russian Federation.
Research of S.V. is supported in part by grant NSh-7989.2016.2 of the President of Russian Federation.
Researches of E.P. and S.V. are supported in part by the RFBR grant 14-01-00707.

\end{document}